\documentclass[fleqn,twoside]{article}
\usepackage{espcrc2}

\usepackage{graphicx,epsfig,color}
\newcommand{\AmS}{{\protect\the\textfont2
  A\kern-.1667em\lower.5ex\hbox{M}\kern-.125emS}}
\title{On the PRISMA Project}

\author{Yuri V. Stenkin
\address{Institute for Nuclear Research of
Russian Academy of Sciences\\
        117312, Moscow, RUSSIA}%
        \thanks{e-mail: stenkin@sci.lebedev.ru}}
\begin{document}

\begin{abstract}
A novel type of Extensive Air Shower (EAS) array is proposed and
described. It is shown that only new approaches to the so called
``knee problem'' could solve this complicated and old problem.
\vspace{1pc}
\end{abstract}
% typeset front matter (including abstract)
\maketitle

\section{Introduction}

There exist (or existed) very few experiments specially designed
to solve the 50 years old ``knee problem'' in cosmic ray. The best
of them, namely, KASCADE and Tibet AS$\gamma$ gave very precise
and interesting but contradicting each other results $\cite{tib}$
and they did not solve the problem. It became even less clear. On
my opinion only new approaches based on the new ideas could solve
this complicated and old problem. The idea of a novel type of
array for EAS study proposed by us for the first time in 2001
$\cite{yus0}$ has been developed in 2008 to the PRISMA (PRImary
Spectrum Measurement Array) project. It is based on a simple idea:
as the hadrons are the main EAS component forming its skeleton and
resulting in all its properties at an observational level
$\cite{Zat}$, then hadron component should also be the main
component to be measured in experiments. Therefore, we have
developed a novel type of EAS array detector ($\it{en-detector}$)
capable to record hadronic component through thermal neutrons
detection and electronic component as well $\cite{yus5}$. The
detector looks like a usual EAS detector but with a specific thin
inorganic scintillator sensitive to thermal neutrons and having
low sensitivity to charged particles. A thin layer of scintillator
consists of an alloy of the mixture of the old inorganic
scintillator ZnS(Ag) plus LiF enriched with $^{6}Li$ up to $90\%$.
Spreading these detectors over a large area on the Earth's surface
one can obtain an hadron calorimeter of practically unlimited
area. Due to rather fast response of the scintillator (the fastest
light component is equal to $\sim40$ ns) these detectors equipped
with constant fraction discriminators can even be used for EAS
timing.

\section{The PRISMA project}
\subsection{Introducing remarks}

As it was already mentioned, the PRISMA experiment is aimed to
solve the ``knee problem'' in cosmic ray spectrum. The best way to
do so could be direct cosmic ray spectrum measurements.
Unfortunately, it can not be performed due to very low intensity
of cosmic ray with energy above 1 PeV. That is why we are pressed
to use an indirect EAS method. But, as a payment for that, one
have to make very complicated and model dependent recalculations
from measured parameters to primary ones. Solving the inverse task
one should be sure that: i) solution exists and ii) measured
parameters are connected with real ones by the known dependencies.
Both points are not known a priori. Solving the direct task one
also have to introduce many parameters by hand, concerning the
using model details, cosmic ray mass composition, existence or
absence of the ``knee'' in primary spectrum etc. $\cite{yus1}$.
Traditionally EAS arrays measure electron component first of all.
This is not the best choice but the simplest and the most
convenient one because the electronic component is the most
numerous one and it produces a great bulk of ionization, which is
used for detection. However, it is the secondary EAS component
that is the mostly sensitive to EAS longitudinal development which
is formed by the cascading high energy hadrons. These two
components are in a dynamic equilibrium. But, the equilibrium
exists only while hadrons exist. When the cascading hadrons are
fully exhausted (note that the number of such hadrons is rather
small below the maximum of shower development and exponentially
decreases up to 0 with the depth in the atmosphere), the
equilibrium violation occurs. This occurs at primary energy of
$\sim$100 TeV/nucleon. It changes the EAS properties dramatically
and results in a visible break (``knee'') in electromagnetic
components (including Cherenkov light as a tertiary component)
$\cite{yus1,yus2}$. Interpretation of the data obtained with
traditional EAS array is very complicated and ambiguous.
Therefore, the best way is to record the primary EAS component,
namely hadronic one. Sure, other components should be record as
well but, mostly for additional and inter-calibrating purposes.
The PRISMA experiment will realize this approach. Similar to a
simple optical prism which splits white light to its components,
the PRISMA will measure EAS in hadronic, muonic and electronic
components separately.

\subsection{Prototypes}

To ensure that proposed idea works properly we constructed two
prototype arrays: one at mountain level (``Multic'', Baksan) and
another one at sea level (``Neutron'', Moscow). Both prototypes
consist now of 4 similar en-detectors. Detectors of ``Multic''
prototype are of 0.375 $m^2$ and that of ``Neutron'' prototype are
of 0.75 $m^2$ each. Detector lay-outs are also different (details
could be found elsewhere $\cite{yus5,yus3,yus4}$). Here only some
preliminary results obtained with the prototypes will be shown and
discussed. Thermal neutron EAS component (``neutron vapor'') could
give us very interesting information that has never been used in
practice before. As it was shown $\cite{yus3}$ these neutrons
accompanying EAS may be of two sources: first one is the
atmosphere ({\it atmospheric neutrons}) and next one is ground or
soil under the array ({\it local neutrons}). These neutrons have
different time distribution due to to different life time in
different media. Life time of thermal neutrons in soil or concrete
or other usual constructed materials (excluding wood) is equal to
$\sim$1 ms, while that in air depends on altitude and is equal to
$\ge$50 ms, in accordance with thermal neutron absorption cross
sections. These two kinds of neutrons carry absolutely different
information about EAS structure and they can be separated
experimentally. Fig.1 shows the results of Monte-Carlo simulations
obtained using CORSIKA (ver. 6501) showers for primary proton and
iron, applied to a prototype setup. It is seen that two branches
of neutrons originated from air and from soil give different time
parameters ($\tau1$ and $\tau2$). The experimental time
distributions obtained with our prototypes $\cite{yus3}$
qualitatively confirm this: both of them can be fitted by a
similar two-exponential curve. In fact the difference between
$\tau1$ and $\tau2$ is not as large as expected due to a mixing
effect. Nevertheless, the difference is enough to be separated
experimentally. The figure shows also that events from primary
protons or iron can be separated using this method.
\begin{figure*}[tbhp]
\vspace{-1.pc} \centerline{\epsfig{height=8cm,file=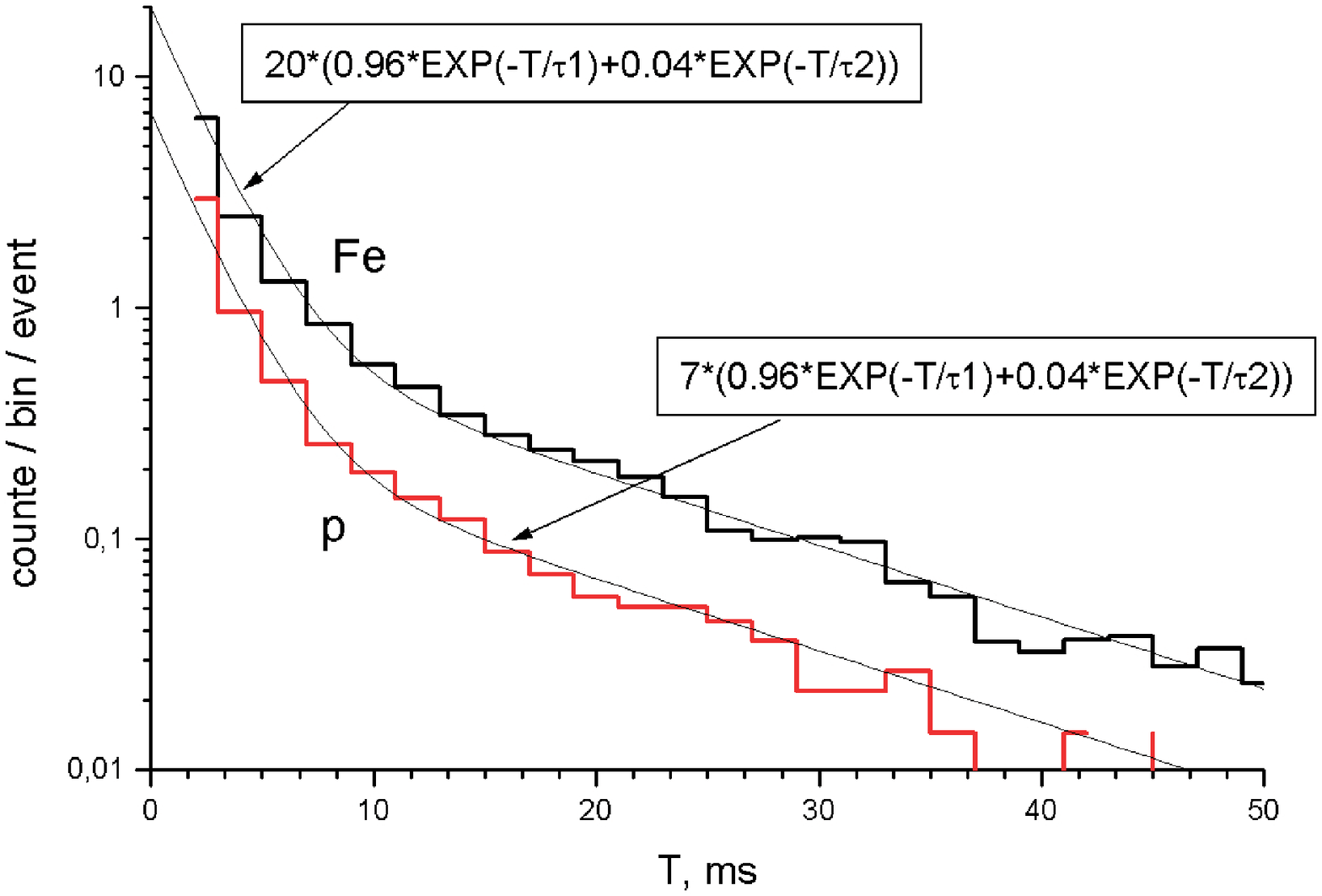}}
\caption{Time distributions. Results of Monte-Carlo simulations %
for primary p and Fe.} \label{fig1}
\centerline{\epsfig{height=9cm,file=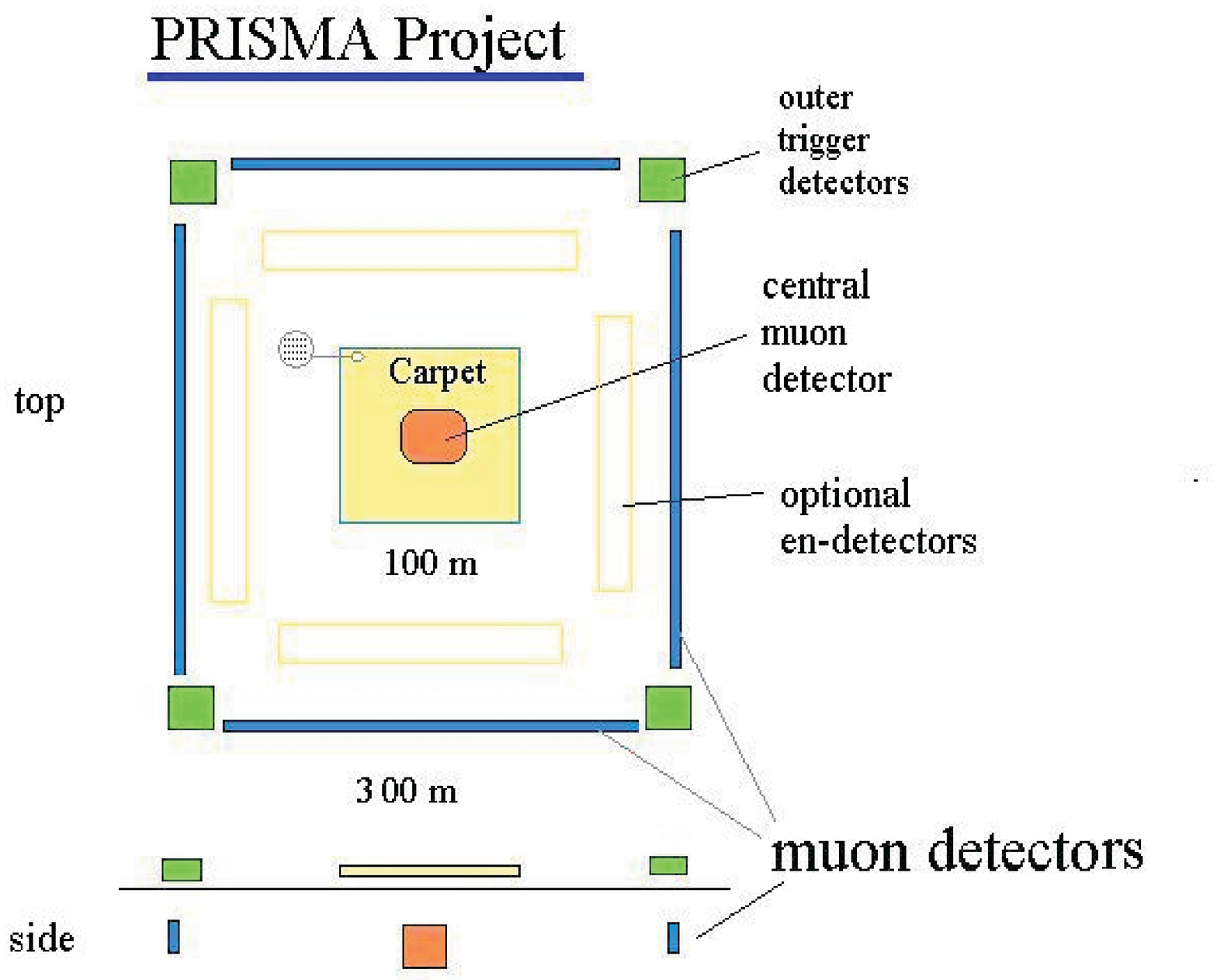}} \caption{The PRISMA
lay-out (top and side view).} \label{fig2}
\end{figure*}
I should emphasize that experimental time distributions could
differ each other being dependent on trigger conditions, on the
array geometry, etc. The higher the shower size, the more neutrons
are detected and different ratio between the two branches is
observed. Sure, the measured time parameters depend on the
experimental details: on the detector distance to EAS core
position, on the media surrounding the detector, array altitude
etc. Our data were obtained for arrays situated inside the
experimental building and should differ from that one could
measure in open air. Nevertheless, the time parameters can be
calculated for any detector location and can be measured
experimentally. In our case we have a difference in time
parameters between two arrays situated at different altitudes,
within a factor of $\sim2$.

\subsection{The PRISMA lay-out.}

Central part of the PRISMA will consist of a large area (at least
100 x 100 $m^2$) covered with en-detectors  $\sim$1 $m^2$ each) as
a rectangle grid with 5 m spacing (see fig. 2). This area is
enough to obtain $\sim10^{4}$ events a year in the ``knee'' region
with cores lies inside it. For higher energy additional outside
en-modules are envisaged. These detectors will record hadronic
(thermal neutrons ``vapor'') and electromagnetic components. This
is one of the project advantages because the same detectors will
record two EAS components and will give two density maps of these
components with a rather good resolution. These maps could be
superimposed and compared off-line. And also, usage of the same
detectors for two purposes makes the project cheaper and more
reliable. The possibility to enlarge the array later without any
problem is an another project advantage. Additional advantages
could be found in $\cite{yus5}$.

Muons are the next important EAS component which give an integral
EAS characteristics. A number of large area muon detectors are
envisaged. They form an outer ring (shaped as a square) consisting
of 1200 individual 1 $m^2$ detectors of the same design as
en-detectors but with usual 5-cm plastic scintillators. Threshold
energy for this detectors is equal to 1 GeV (under 500 $g/cm^{2}$
of soil absorber). The central underground muon detector design is
not still fixed. Probably it could be a fine-structured track
detector of at least 100 $m^2$ in total or it could look like a
continuous carpet of 20x20 individual 1 $m^2$ detectors.  And
finally, 4 x 25 such detectors placed on the surface will be used
as outer trigger detectors.

\begin{table}[htbp!]
\caption{Main features of the PRISMA array.}
\label{table:1}
\newcommand{\m}{\hphantom{$-$}}
\newcommand{\cc}[1]{\multicolumn{1}{c}{#1}}
\renewcommand{\tabcolsep}{1pc} % enlarge column spacing
\renewcommand{\arraystretch}{1.3} % enlarge line spacing
\begin{tabular}{@{}llll}
\hline
primary energy range, (eV)           & \m $\sim 10^{13} - 10^{16}$\\
energy resolution, (\%)              & \m $\sim 10$  \\
angular resolution, (degree)         & \m $\sim1$    \\
core location accuracy, (m)          & \m $<2$     \\
\hline
\end{tabular}\\[2pt]
\end{table}
\section{Conclusion}

The project of a novel type of EAS array is proposed. We do
believe when running this array will solve the ``knee'' problem.
Location of this experiment is not fixed yet. It depends on the
collaboration of institutions, which is still open for other
participants. High altitude location is more preferable. It would
be very interesting to locate PRISMA at the Tibet high mountain
plateau nearby the existing arrays of Tibet AS$\gamma$  and Argo
YBJ or combine it with recently proposed $\cite{hua}$ new project LHAASO. Any
new proposals and collaborators are welcome.

This work was partially supported by RFBR grants 08-02-01208,
07-02-00964 and by the ``Neutrino program'' of Russian Academy of
Sciences.

\end{document}